\documentclass[conference]{IEEEtran}
\IEEEoverridecommandlockouts

\usepackage{graphicx}
\usepackage{tikz}
\graphicspath{{./}}

\usepackage{amsmath}
\usepackage{amssymb}
\usepackage{amsfonts}
\usepackage{cases}

\usepackage{float} 
\newfloat{footnote}{hb} 

\usepackage{acronym}

\usepackage{color}
\usepackage{multicol}
\usepackage{stfloats}
\usepackage{hyperref}
\usepackage{caption}
\usepackage{subcaption}

\usepackage{balance}

\usepackage{subfiles}
 
\usepackage{setspace}
\usepackage{lipsum}

\usepackage{amssymb, textcomp}
\usepackage{amsfonts}
\usepackage{color}
\usepackage{amssymb, latexsym, amsmath, bm, dcolumn}
\usepackage{algpseudocode}
\usepackage{amsmath, color, epsf}
\usepackage[linesnumbered,ruled,vlined]{algorithm2e}

\DeclareMathAlphabet\mathbfcal{OMS}{cmsy}{b}{n}
\begin{document}
\title{An Accelerated Stochastic Gradient for Canonical Polyadic Decomposition\\
\thanks{ All members were supported by the European Regional Development Fund of the European Union and Greek national funds through the Operational Program Competitiveness, Entrepreneurship, and Innovation, under the call RESEARCH - CREATE - INNOVATE (project code : ${\rm T1E\Delta K-03360}$).}}
%

\author{\IEEEauthorblockN{Ioanna Siaminou}
\IEEEauthorblockA{\textit{School of Electrical and Computer Engineering} \\
\textit{Technical University of Crete}\\
		Chania, Greece \\
		isiaminou@isc.tuc.gr}
\and
\IEEEauthorblockN{Athanasios P. Liavas}
\IEEEauthorblockA{\textit{School of Electrical and Computer  Engineering} \\
	\textit{Technical University of Crete}\\
	Chania, Greece \\
	liavas@telecom.tuc.gr}
}
\maketitle
\begin{abstract}
We consider the problem of structured canonical polyadic decomposition.
If the size of the problem is very big, then stochastic gradient approaches are viable alternatives to
classical methods, such as Alternating Optimization and All-At-Once optimization.
We extend a recent stochastic gradient approach by employing 
an acceleration step (Nesterov momentum) in each iteration. 
We compare our approach with state-of-the-art alternatives, using both synthetic and real-world data, and find it to be very competitive.
\end{abstract}

\section{Introduction}
\label{sec:intro}

The need to understand and analyze multi-dimensional data and their interdependencies has led to the extended use of tensors in diverse scientific fields, ranging, for example, from medicine to geodesy. 
An overview of tensor applications can be found in \cite{rabanser2017introduction}. \cite{sidiropoulos2017tensor}.

Canonical Polyadic Decomposition (CPD) or Parallel Factor Analysis (PARAFAC) is a widely used model  since, in many cases, it extracts meaningful structure from a given dataset. 

Alternating Optimization (AO), All-at-Once Optimization(AOO), and Multiplicative Updates (MUs) are the workhorse methods towards the computation of the CPD \cite{cichocki2009nonnegative}, \cite{CichockiMPCZZL14}. 
However, the very large size of the collected tensor data makes the implementation of these
algorithms very computationally demanding.

Recently, various approaches have been introduced in order to deal with large-scale CPD problems. 
An obvious approach is the development and implementation of parallel algorithms (distributed or shared-memory) \cite{Ballard_et_al_2018}, \cite{smith2016medium}, \cite{liavas2017nesterov}. 

From a different perspective,  stochastic gradient based algorithms have gained much attention, since they are relatively easy to implement, have lower computational cost, and can guarantee accurate solutions.

\subsection{Related Work}
\label{subsec:Related_Work}
Sub-sampling of the target tensor $\mathbfcal{X}$ using regular sampling techniques has been introduced in \cite{8902959}. 

In \cite{vervliet2015randomized}, entries of the target tensor are sampled in a random manner and the respective blocks of the latent factors are updated at each iteration. 

In \cite{papalexakis2012parcube} and \cite{sidiropoulos2014parallel}, a distributed framework is employed, where smaller replicas of the target tensor are independently factored. The resulting factors of each independent decomposition are effectively merged at the end to obtain the final latent factors. 

In \cite{battaglino2018practical} and \cite{fu2020block}, a set of fibers is randomly selected at each iteration and a stochastic proximal gradient step is performed. 
We improve upon the work of \cite{fu2020block} by incorporating Nesterov acceleration at each iteration and a proximal term to deal with ill-conditioned cases.

\subsection{Notation}
\label{subsec:Notation}

Vectors and matrices are denoted by lowercase and uppercase bold letters, for example, $\mathbf{x}$ and $\mathbf{X}$. Tensors are denoted by bold calligraphic capital letters, namely, $\mathbf{\mathbfcal{X}}$. $\| \cdot \|_F$ denotes the Frobenius norm of the matrix or tensor argument. $\mathbf{A} \odot \mathbf{B}$ and $\mathbf{A} \circledast \mathbf{B}$ denote, respectively, the Khatri-Rao and the Hadamard product of matrices ${\bf A}$ and ${\bf B}$. The outer product between vectors is denoted by the operator $\circ$. Finally, MATLAB notation is used when it seems appropriate.

\section{Canonical Polyadic Decomposition}
\label{sec:cpd}

Let tensor $\mathbfcal{X}^o \in \mathbb{R}^{I_1 \times I_2 \times \dots \times I_N}$ admit a 
rank-$R$ factorization of the form
\begin{equation}
\mathbfcal{X}^o \hspace{-0.1cm}= \mbox{\textlbrackdbl} \mathbf{A}^{o(1)}, \dots, \mathbf{A}^{o(N)}
\mbox{\textrbrackdbl} =
\sum_{r=1}^R \mathbf{a}_r^{o(1)}\hspace{-.1cm}\circ \dots \circ \mathbf{a}_r^{o(N)},
\label{CPD_def}
\end{equation}
where $\mathbf{A}^{o(i)}=[\mathbf{a}_1^{o(i)} ~\cdots ~ \mathbf{a}_R^{o(i)}]\in\mathbb{R}^{I_i \times 
R}$, for $i =1,\ldots,N$.
In many cases, the latent factors have a special structure or satisfy a specific property, which is denoted as
${\bf A}^{o(i)}\in \mathbb{A}^i$. 

The actual observation is corrupted by additive noise $\mathbfcal{E}$,
thus, we observe tensor $\mathbfcal{X} = \mathbfcal{X}^o + \mathbfcal{E}$.

Estimates of $\mathbf{A}^{o(i)}$ can be obtained by
computing matrices $\mathbf{A}^{(i)}$ that solve the optimization problem
\begin{equation}
\label{LS_with_con}
 \underset {\{ \mathbf{A}^{(i)} \in\mathbb{A}^i \}_{i=1}^n } \min 
f_{\mathbfcal{X}}\left(\mathbf{A}^{(1)}, \dots , \mathbf{A}^{(N)}\right),
\end{equation}
where $f_{\mathbfcal{X}}$ is a function that measures the quality of the 
factorization, with a common choice being
\begin{equation}
f_{\mathbfcal{X}}\left(\mathbf{A}^{(1)}, \dots , \mathbf{A}^{(N)}\right) 
=  \left\| \mathbfcal{X} - \mbox{\textlbrackdbl} \mathbf{A}^{(1)},  \dots, \mathbf{A}^{(N)}
\mbox{\textrbrackdbl} \right\|_F^2\hspace{-0.1cm}.
\label{CPD_problem}
\end{equation}
This optimization problem is nonconvex and, thus, difficult to solve.
The matrix unfolding (or tensor matricization) has been very useful towards the solution 
of problem (\ref{CPD_problem}).
More specifically, 
if $\mathbfcal{\widehat{X}}= \mbox{\textlbrackdbl} \mathbf{A}^{(1)}, \dots, \mathbf{A}^{(N)} \mbox{\textrbrackdbl}$, 
then the matrix unfolding for an arbitrary mode $i$  is given by  \cite{KoBa09}, \cite{sidiropoulos2017tensor}
\vspace{1em}
\begin{equation}
{\bf \widehat{X}}^{(i)}  =  \mathbf{K}^{(i)} \mathbf{A}^{(i)T},
\label{mt_approximate_factors}
\end{equation}
\vspace{1em}
where ${\bf K}^{(i)}$ is defined as
\begin{equation}
\label{K_KRPR}
\mathbf{K}^{(i)} := \mathbf{A}^{(N)} \hspace{-0.1cm} \odot \dots \odot \hspace{-0.06cm}  \mathbf{A}^{(i+1)} \hspace{-0.1cm}  \odot \mathbf{A}^{(i-1)} \hspace{-0.1cm}  \odot \dots \odot 
\mathbf{A}^{(1)}.
\end{equation}
It can be shown that, for $i=1,\ldots,N$,
\begin{equation}
\begin{split}
f_{\mathbfcal{X}}(\mathbf{A}^{(1)}, \dots , \mathbf{A}^{(N)}) & = 
\,\left\| \mathbf{X}^{(i)} - \mathbf{\widehat{X}}^{(i)} \right\|_F^2.
\end{split}
\label{f_X_matr}
\end{equation}
These expressions form the basis of the CPD AO method. 
If, at iteration $k$, the estimated
matrix factors have values $\mathbf{A}_k^{(j)}$, for $j=1,\ldots,N$, 
we can update $\mathbf{A}_k^{(i)}$ 
by solving the matrix least-squares problem
\begin{equation}
\label{LS}
\mathbf{A}_{k+1}^{(i)} = ~ \underset{\mathbf{A}^{(i)}\in\mathbb{A}^i}{\rm argmin} ~  \| \mathbf{X}^{(i)} - \mathbf{K}_k^{(i)}
\mathbf{A}^{(i)T} \|_F^2,
\end{equation}
where
\begin{equation}
\label{K_KRPR}
\mathbf{K}_k^{(i)} := \mathbf{A}_k^{(N)} \hspace{-0.1cm} \odot \dots \odot \hspace{-0.06cm}  \mathbf{A}_k^{(i+1)} \hspace{-0.1cm}  \odot \mathbf{A}_{k+1}^{(i-1)} \hspace{-0.1cm}  \odot \dots \odot 
\mathbf{A}_{k+1}^{(1)}.
\end{equation}
The gradient of $f_{\mathbfcal X}$,
with respect to ${\bf A}^{(i)}$, is given by
\begin{equation}
\begin{split}
& \nabla_{{\bf A}^{(i)}} f_{\mathbfcal{X}}({\bf A}_k^{(1)},{\bf A}_k^{(2)}, \ldots, {\bf A}_k^{(N)}) \cr
& \quad \qquad = {\bf A}_k^{(i)}  {\bf K}_k^{(i)T} {\bf K}_k^{(i)} - {\bf X}^{(i)T} {\bf K}_k^{(i)}.
\end{split}
\end{equation}
Quantity ${\bf X}^{(i)T} {\bf K}_k^{(i)}$ is called Matricized Tensor Times Khatri-Rao Product (MTTKRP) and is the most computationally demanding part of
the CPD AO algorithm. Thus, the development of efficient algorithms which do not compute a full MTTKRP during each iteration is of great interest. 

\section{Stochastic Gradient CPD - BrasCPD}
\label{sec:BrasCPDaccel}

In \cite{vervliet2015randomized}, a stochastic gradient method has been developed for the unconstrained case, where, at each iteration, only a part of a factor is updated. 

A fiber sampling technique has been developed in \cite{battaglino2018practical} and \cite{fu2020block}, where, at each iteration, a whole factor is updated. The scheme proposed in \cite{fu2020block}, named BrasCPD, can handle both unconstrained and constrained problems.

BrasCPD combines the fiber sampling technique with the AO algorithm. 
Assume, again, that, at the beginning of iteration $k$, the values of the estimated factors are ${\bf A}_k^{(j)}$, for
$j=1,\ldots,N$. 
At iteration $k$, an index $i$ is picked at random. Then, $B^i$ mode-$i$ fibers are sampled, indexed by $\mathcal{F}_k^i \subset \lbrace 1,2, \dots J^i \rbrace$, where $J^i$ denotes the number of rows of ${\bf  X}^{(i)}$, and a smaller problem is
constructed, namely,
\begin{equation}
\label{LS_small}
\underset{{\bf A}^{(i)}\in\mathbb{A}^i}{\min}
f_k^{(i)}({\bf A}^{(i)}), 
\end{equation}
where
\begin{equation*}
f_k^{(i)}({\bf A}^{(i)}) =
\| {\bf X}^{(i)}(\mathcal{F}_k^i,:) - {\bf K}_k^{(i)}(\mathcal{F}_k^i,:){\bf A}^{(i)T} \|_F^2.
\end{equation*}
BrasCPD performs a proximal gradient step and updates ${\bf A}^{(i)}_k$ as
\begin{equation}
{\bf A}^{(i)}_{k+1}  = {\rm prox}_{h_i} \left( {\bf A}^{(i)}_{k} - \frac{\alpha_{k}}{|{\cal F}_k^i|}
\nabla f_k^{(i)}({\bf A})
 \right), 
\end{equation}
where $h_i$ is the indicator function of set $\mathbb{A}^i$ and 
\begin{equation}
\begin{split}
\nabla f_k^{(i)}({\bf A}) & =  {\bf A}^{(i)}_k{\bf K}_k^{(i)T}(\mathcal{F}_k^i,:){\bf K}_k^{(i)}(\mathcal{F}_k^i,:) \cr
& \quad\quad - {\bf X}^{(i)T}(\mathcal{F}_k^i,:){\bf K}_k^{(i)}(\mathcal{F}_k^i,:).
\end{split}
\label{nabla_f_k_i}
\end{equation}
The other factors do not change during iteration $k$, that is, for $j\ne i$,
${\bf A}^{(j)}_{k+1} = {\bf A}^{(j)}_{k}$.
Note that the computational cost of the {\em partial}\/ MTTKRP appearing in (\ref{nabla_f_k_i}) 
drops to $\mathcal{O}(|\mathcal{F}_k^i| R I_i )$ flops.

The performance of the algorithm is mainly determined by the step-sizes $\alpha_k$ \cite{bottou2018optimization}. 
In BrasCPD, diminishing step-sizes are employed, namely,
$\alpha_k = \frac{\alpha}{k^{\beta}}$,
for appropriate values of parameters $\alpha$ and $\beta$.

A method based on Adagrad \cite{duchi2011adaptive}, called AdaCPD, has also been proposed in \cite{fu2020block}. 
The AdaCPD computes its step-sizes using an accumulated-gradient mechanism with
parameters $\eta > 0$, $\beta > 0$,  and $\epsilon>0$, namely 
\vspace{1em}
\begin{equation}
\alpha_k^{(i)} = \frac{\eta}{\left( \beta + \sum_{l=1}^{k}{\nabla f_l^{(i)}({\bf A})}\right)^{1/2 + \epsilon}}.
\end{equation}

\section{Accelerated Stochastic Gradient CPD}
\label{section_ASCPD}

We propose an accelerated version of BrasCPD, which we call  Accelerated Stochastic CPD (ASCPD). 
At iteration $k$, we follow the same sampling scheme as in \cite{fu2020block}, \cite{battaglino2018practical},
and form the problem 
\begin{equation}
\label{LS_small_NAG}
\underset{{\bf A}^{(i)}\in\mathbb{A}^i}{\rm min}  
F_k^{(i)}({\bf A}^{(i)}) = 
f_k^{(i)}({\bf A}^{(i)})
+ \frac{\lambda^{(i)}_k}{2} \| {\bf A}^{(i)} - {\bf A}^{(i)}_k\|_F^2.
\end{equation}
The parameter $\lambda^{(i)}_k$ determines the condition number of the problem and is chosen
such that the condition number remains ``small.'' More specifically, the Hessian of
$f_k^{(i)}$ is 
\vspace{1em}
\begin{equation}
\hspace{-.1cm}
{\bf H}_k^{(i)}\hspace{-.1cm}:=
\hspace{-.05cm}\nabla^2  f_k^{(i)}({\bf A}^{(i)})  = 
{\bf K}_k^{(i)T}({\cal F}^i_k,:) {\bf K}_k^{(i)}({\cal F}^i_k,:) \otimes {\bf I}_{I_iR}.
\label{Hessian_f_k_i}
\end{equation}
Let $L_k^{(i)}$ and $\mu_k^{(i)}$ be, respectively, the largest and 
smallest eigenvalues of ${\bf H}_k^{(i)}$. We choose $\lambda_k^{(i)}$ such that
the condition number of (\ref{LS_small_NAG}) becomes ``almost constant,'' that is
\vspace{1em}
\begin{equation}
\frac{\bar{L}^{(i)}_k}{\bar{\mu}^{(i)}_k}:=
\frac{L_k^{(i)}+\lambda_k^{(i)}}{\mu_k^{(i)}+\lambda_k^{(i)}} \lesssim {\cal C},
\end{equation}
\vspace{1em}
where ${\cal C}$ is a given value (in our experiments, we set ${\cal C}=10, 10^2, 10^3$).
More specifically, we set
\begin{equation}
\lambda_k^{(i)}=\left\{ \begin{array}{ll} 
\mu_k^{(i)}, & \mbox{if}~\frac{L_k^{(i)}}{\mu_k^{(i)}} < {\cal C}, \cr
\frac{L_k^{(i)}}{\cal C}, & \mbox{otherwise}.
\end{array}
\right.
\end{equation}
We perform a proximal step
\begin{equation}
\label{update_NAG}
{\bf A}^{(i)}_{k+1} = {\rm prox}_{h_i} 
\left( {\bf Y}^{(i)}_{k} - \frac{1}{\bar{L}^{(i)}_k}
\nabla F_k^{(i)}({\bf Y}_k^{(i)}) \right),
\end{equation}
followed by a momentum step
\begin{equation}
\label{interpolation_NAG}
{\bf Y}_{k+1}^{(i)} = {\bf A}_{k+1}^{(i)} + \beta_k^{(i)}
\left( {\bf  A}_{k+1}^{(i)} - {\bf A}_k^{(i)} \right),
\end{equation}
where
\begin{equation}
\beta_k^{(i)} := \frac{ \sqrt{\bar{L}_k^{(i)}}-\sqrt{\bar{\mu}_k^{(i)}} }
{\sqrt{\bar{L}_k^{(i)}}+\sqrt{\bar{\mu}_k^{(i)}}}.
\end{equation}
The values of $\bar{L}^{(i)}_k$ and $\beta_k^{(i)}$ are the steps used by the ``constant step scheme III'' of the
accelerated gradient algorithm \cite[p. ~81]{Nesterov_Book_2004}. They can be considered
as ``locally optimal''  for problem (\ref{LS_small_NAG}). 

Note that we essentially compute  
${\bf H}_k^{(i)}$ during the computation of $\nabla f_k^{(i)}$ (see (\ref{nabla_f_k_i}) and (\ref{Hessian_f_k_i})). Furthermore, the computation
of its largest and smallest eigenvalues does not pose significant computational cost, especially
in the cases of small $R$. 

The ASCPD algorithm appears in Algorithm \ref{BrasCPDaccel_algo}.
A variation of our scheme is to use only the stochastic proximal gradient step of (\ref{update_NAG}), without the acceleration step. We will test the effectiveness of this variation in our
numerical experiments. 

An important future research topic is the development of algorithms that fully exploit the second-order information ${\bf H}_k^{(i)}$. Some initial efforts have not resulted in algorithms superior to the one proposed in this paper, especially in the noisy cases.

\begin{algorithm}[h]
\SetAlgoLined
\KwResult{ $\lbrace {\bf  A}^{(i)} \rbrace_{i=1}^N$}
 \textbf{Input:}{$~ {\rm tensor}~ \mathbfcal{X}, 
 {\bf A}^{(i)}_{0}={\bf Y}^{(i)}_{0}$, $i=1,\ldots,N$, 
 blocksizes $B^i$, $i=1,\ldots,N$.
 \\$\textit{k} = 0$}\;
 \While{terminating condition is not satisfied}{
 {
   Uniformly sample $i$ from $\lbrace 1,2 \dots N \rbrace$ \;
   Uniformly sample $B^i$ mode-$i$ fibers\;
   Compute stochastic gradient $\nabla F^{(i)}_k({\bf Y}_k^{(i)})$\;
   Compute $L_k^{(i)}$, $\mu_k^{(i)}$, and $\lambda_k^{(i)}$\;
   Compute ${\bf A}^{(i)}_{k+1}$ using (\ref{update_NAG}) \;
   Compute ${\bf Y}^{(i)}_{k+1}$ using (\ref{interpolation_NAG}) \;
   \textit{k} = \textit{k} + 1\;
  }
 }
 \caption{ASCPD}
 \label{BrasCPDaccel_algo}
\end{algorithm}

\section{Numerical Experiments}
\label{sec:num_exps}

In this section, we run numerical experiments and test the performance, in terms of convergence speed and estimation accuracy,  of the NALS algorithm of \cite{liavas2017nesterov}, AdaCPD, ASCPD, and 
BrasCPD with locally optimal step-size, using both synthetic and real-world data.

In our experiments, 
AdaCPD always outperformed BrasCPD, thus, we do not consider BrasCPD. For both synthetic and real-world data, the results are extracted after averaging over $10$ Monte-Carlo trials.

\subsection{Synthetic Data}

\begin{figure}
\centerline{\includegraphics[width=9.5cm]{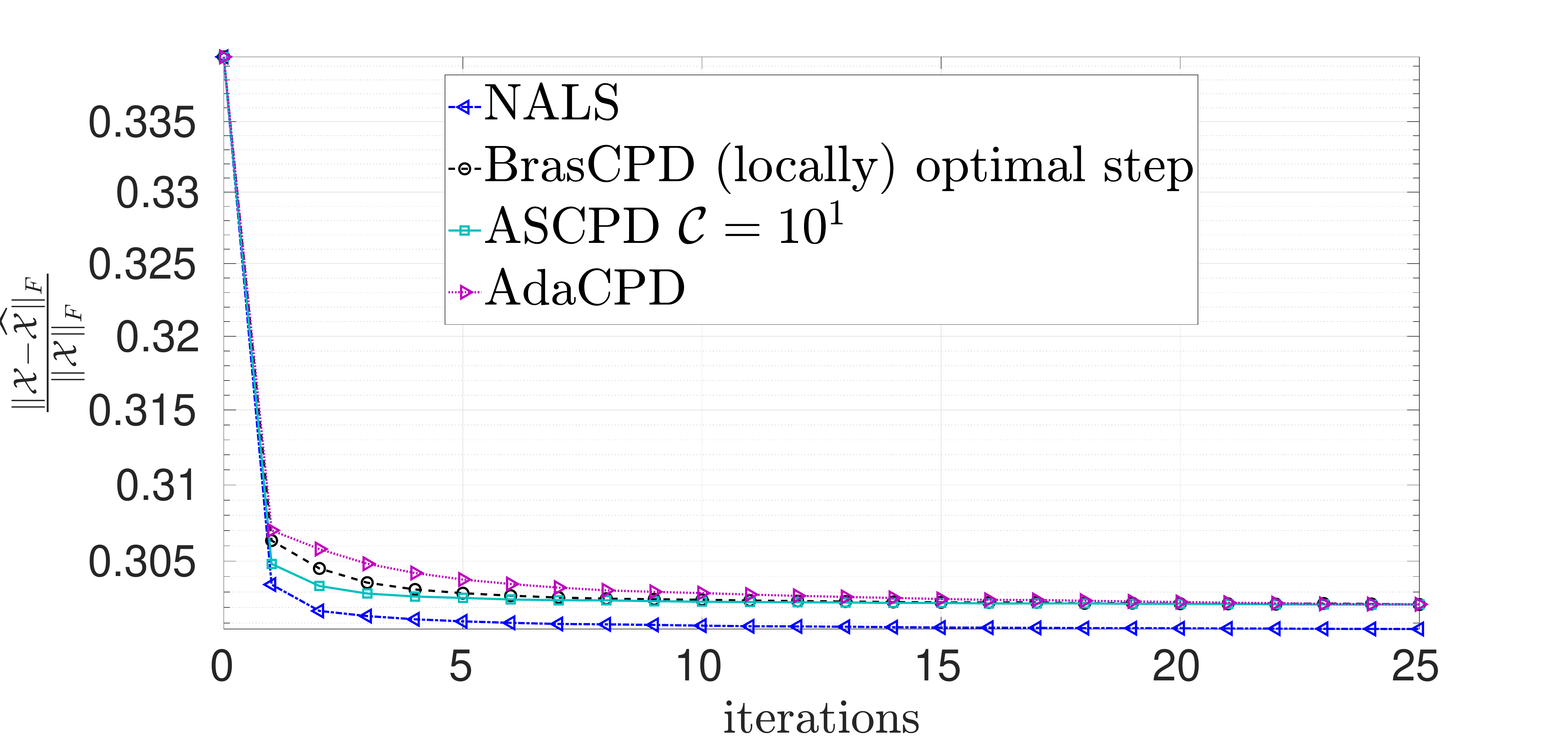}}
\centerline{\includegraphics[width=9.5cm]{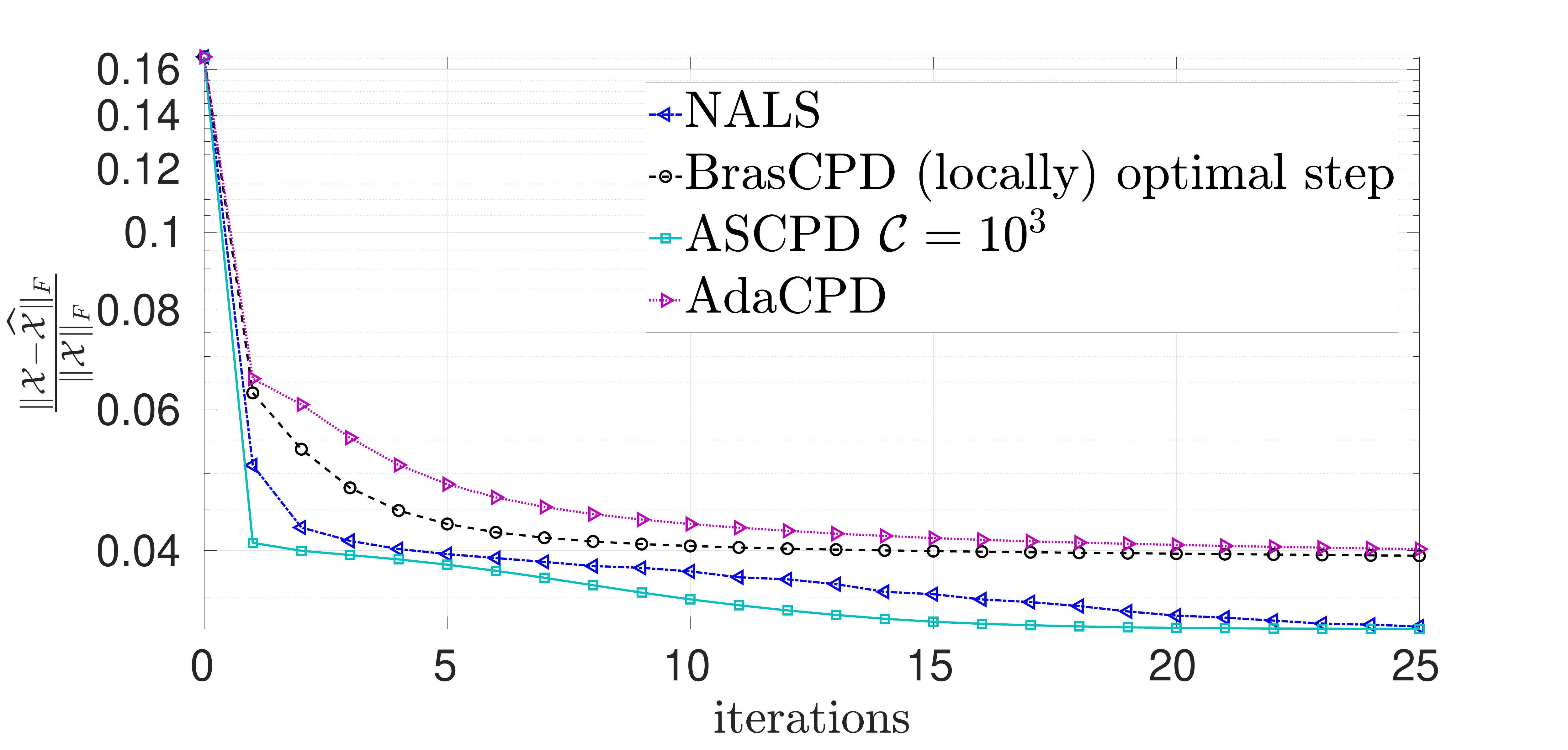}}
\caption{Relative tensor reconstruction error for $I_1=I_2=I_3=200$, $R=100$, 
$B=500$, and ${\tt SNR}= 10{\rm dB}$  (top) and $30$\,dB (bottom).}
\label{fig_I_200_R_100_B_500_SNR_10_30}
\end{figure}

\begin{figure}
\centerline{\includegraphics[width=9.5cm]{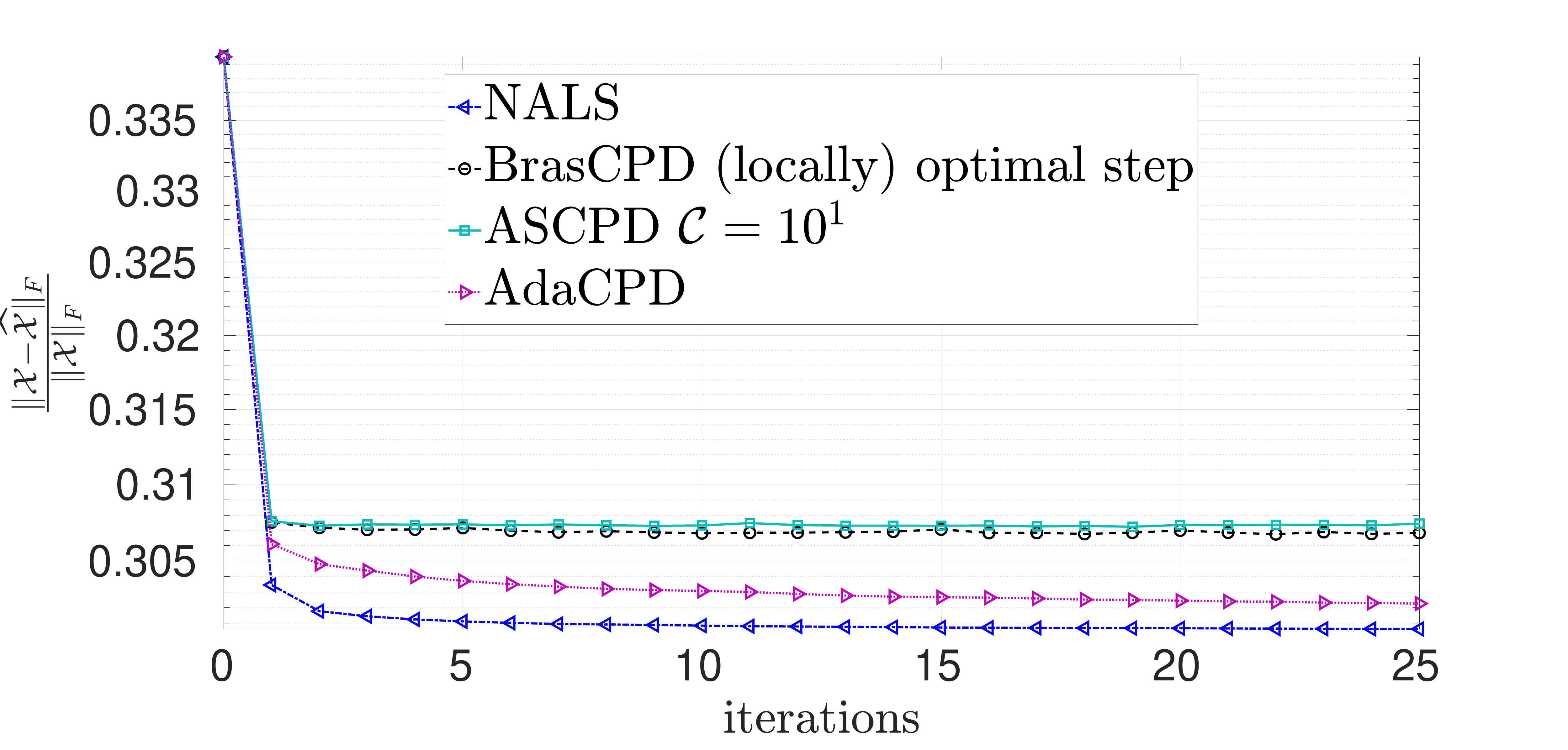}}
\centerline{\includegraphics[width=9.5cm]{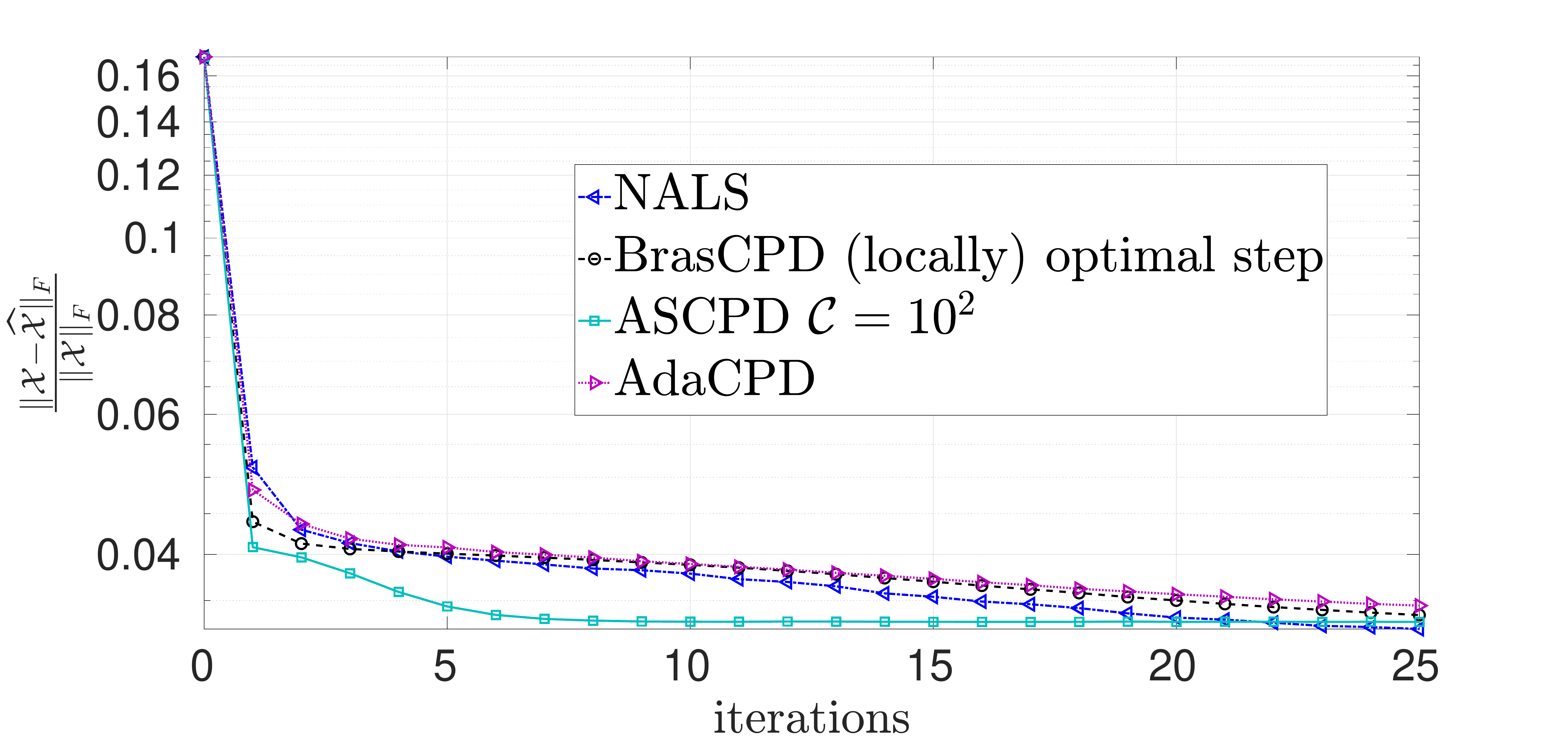}}
\caption{Relative tensor reconstruction error for $I_1=I_2=I_3=200$, 
$R=100$, $B=100$, and ${\tt SNR} = 10$\,dB (top) and $30$\,dB (bottom).}
\label{fig_I_200_R_100_B_100_SNR_10_30}
\end{figure}

We generate a $3$-rd order nonnegative tensor $\mathbfcal{X}^o \in \mathbb{R}_+^{I_1 \times I_2 \times I_3}$ as
$\mathbfcal{X}^o = \mbox{\textlbrackdbl} \mathbf{A}^{o(1)}, \dots, \mathbf{A}^{o(N)}
\mbox{\textrbrackdbl}$, 
where the elements of each factor are independent and identically distributed (i.i.d.), uniformly distributed in $[0,1]$.
The noisy tensor is given by
$\mathbfcal{X} = \mathbfcal{X}^o + \sigma_{\epsilon} \mathbfcal{E}$, 
where the elements of $\mathbfcal{E}$ are i.i.d. ${\cal N}(0,1)$.
 We define the Signal-to-Noise Ratio (SNR) 
\begin{equation*}
{\tt SNR} := \frac{\| \mathbfcal{X}^o \|^2_F}{\sigma_{\epsilon}^2 \| \mathbfcal{E} \|^2_F}.
\end{equation*}
We adopt as performance metric the relative tensor reconstruction error at iteration $k$
\begin{equation*}
m_k:= \frac{\| \mathbfcal{X} - \mathbfcal{\widehat{X}}_k\|_F}{\| \mathbfcal{X} \|_F}.
\end{equation*}
All stochastic gradient based algorithms use the same blocksize, that is, $B^1=B^2=B^3=B$. 
Furthermore, we note that one full iteration of the NALS algorithm of \cite{liavas2017nesterov}
requires the computation of
four full MTTKRPs (three for the factor updates and one for the acceleration step).
Thus, in order to be fair,
in our plots, we depict the performance metric $m_k$ attained by each algorithm after the 
{\em same number of full iterations}, that is, we compute the performance metric for each stochastic algorithm after the required number of stochastic iterations that correspond to one full iteration of the NALS algorithm.

In Fig. \ref{fig_I_200_R_100_B_500_SNR_10_30}, we plot the metric $m_k$ versus the number of full iterations for the case
where $I_1=I_2=I_3=200$, $R=100$, $B=500$, and SNR = $10$\,dB (top) and $30$\,dB (bottom).

In Fig. \ref{fig_I_200_R_100_B_100_SNR_10_30}, we set $B=100$ and present the corresponding plot. 
In all cases, we set the Adagrad parameter $\eta=1$ (in our experiments, this was the best value), while we set the ASCPD parameter ${\cal C}= 10$ 
in the low SNR cases and ${\cal C}= 10^2, 10^3$ in the high SNR cases.

We observe that
\begin{enumerate}
\itemsep0em
\item in the low SNR cases, the NALS algorithm outperforms all stochastic gradient based approaches.
 The relative performance of AdaCPD and ASCPD depends on the block size. For ``small'' block sizes, AdaCPD outperforms the ASCPD, while, for ``large'' block sizes, the opposite happens. 
\item 
in the high SNR cases, the ASCPD outperforms all other methods. We note that the BrasCPD with
locally optimal step-size in some cases outperforms AdaCPD.
\end{enumerate}

\begin{figure}
\centerline{\includegraphics[width=9.5cm]{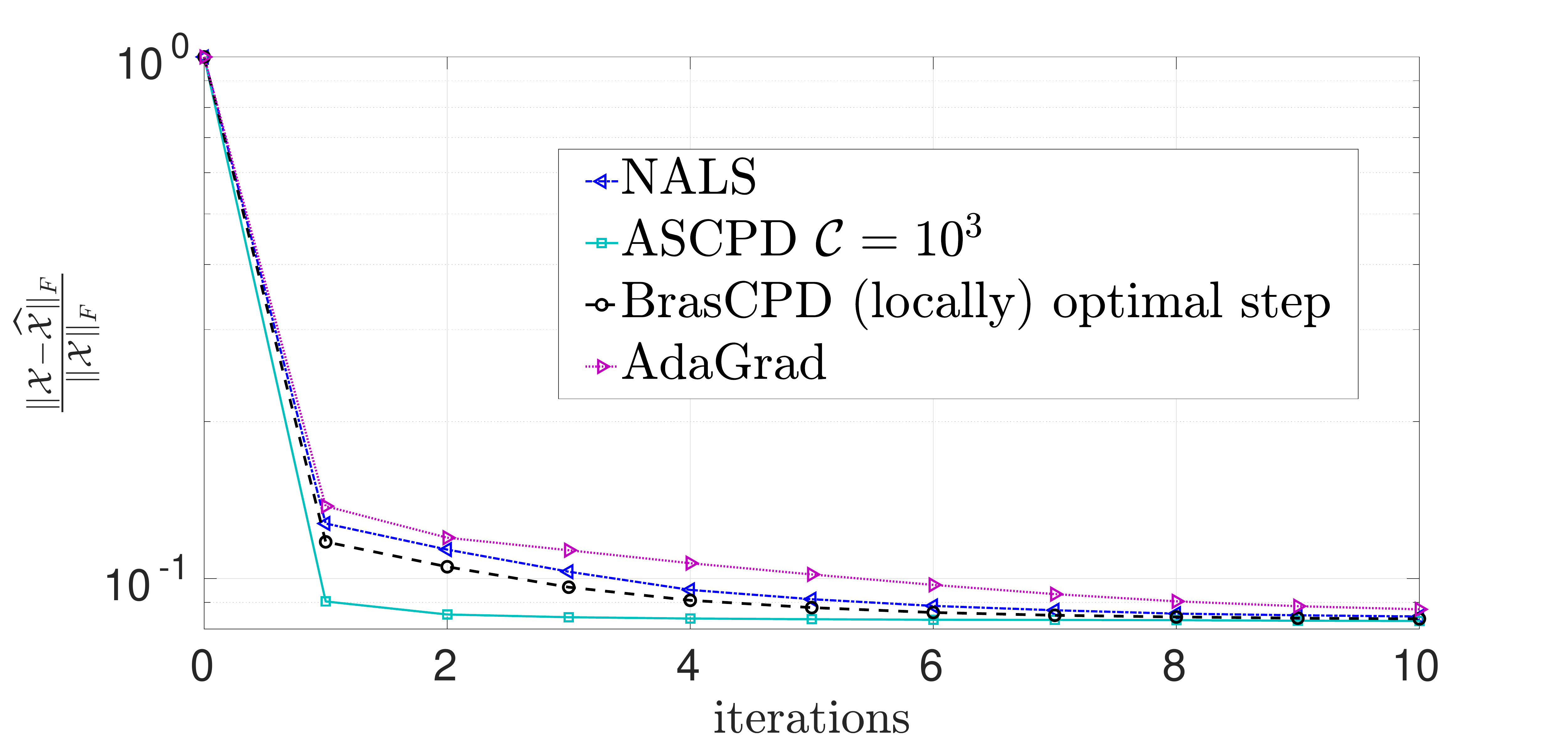}}
\centerline{\includegraphics[width=9.5cm]{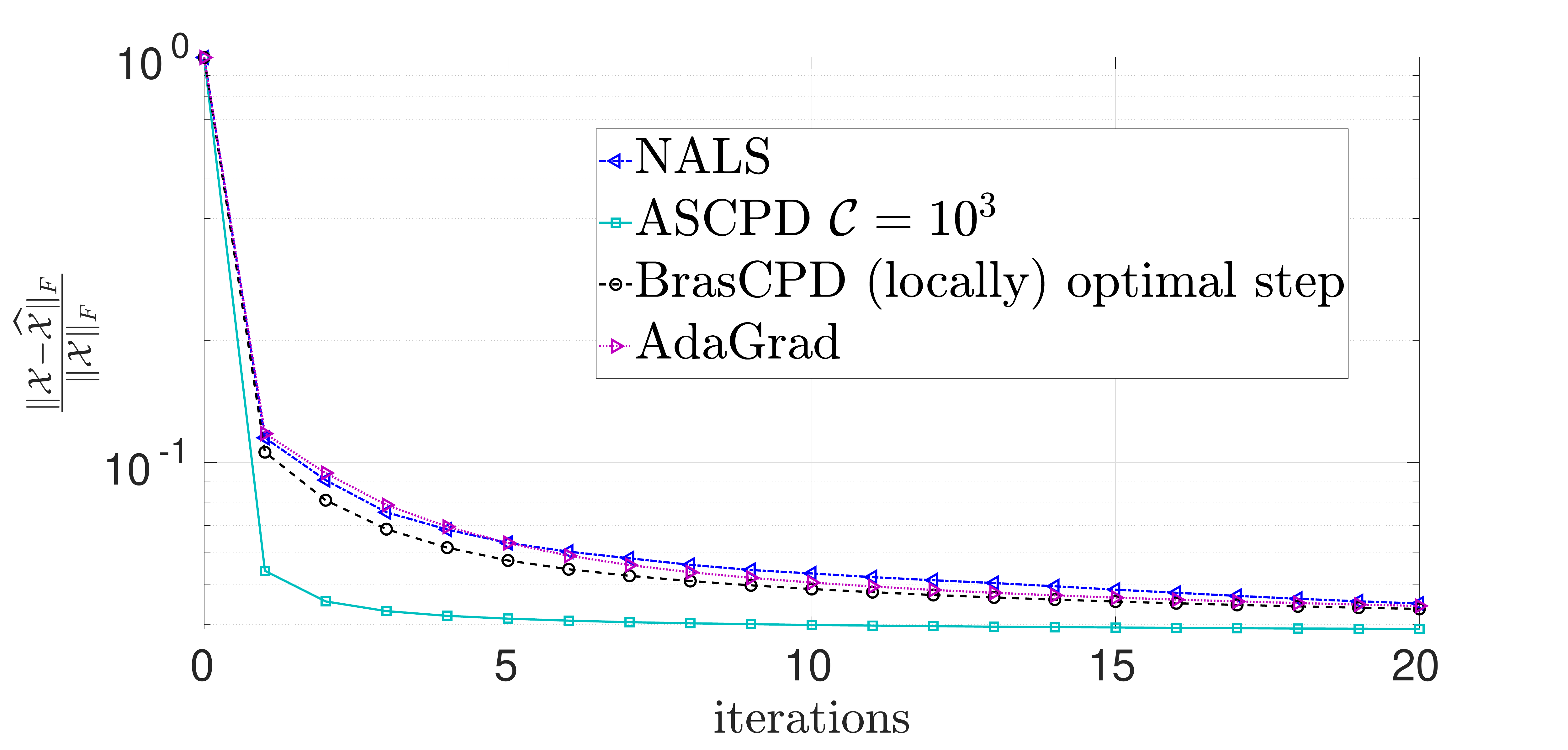}}
\caption{Indian Pines dataset: relative tensor reconstruction error
for $B=500$, and $R=10$ (top), $R = 100$ (bottom).}
\label{fig_indianpine}
\end{figure}

\begin{figure}
\centerline{\includegraphics[width=9.5cm]{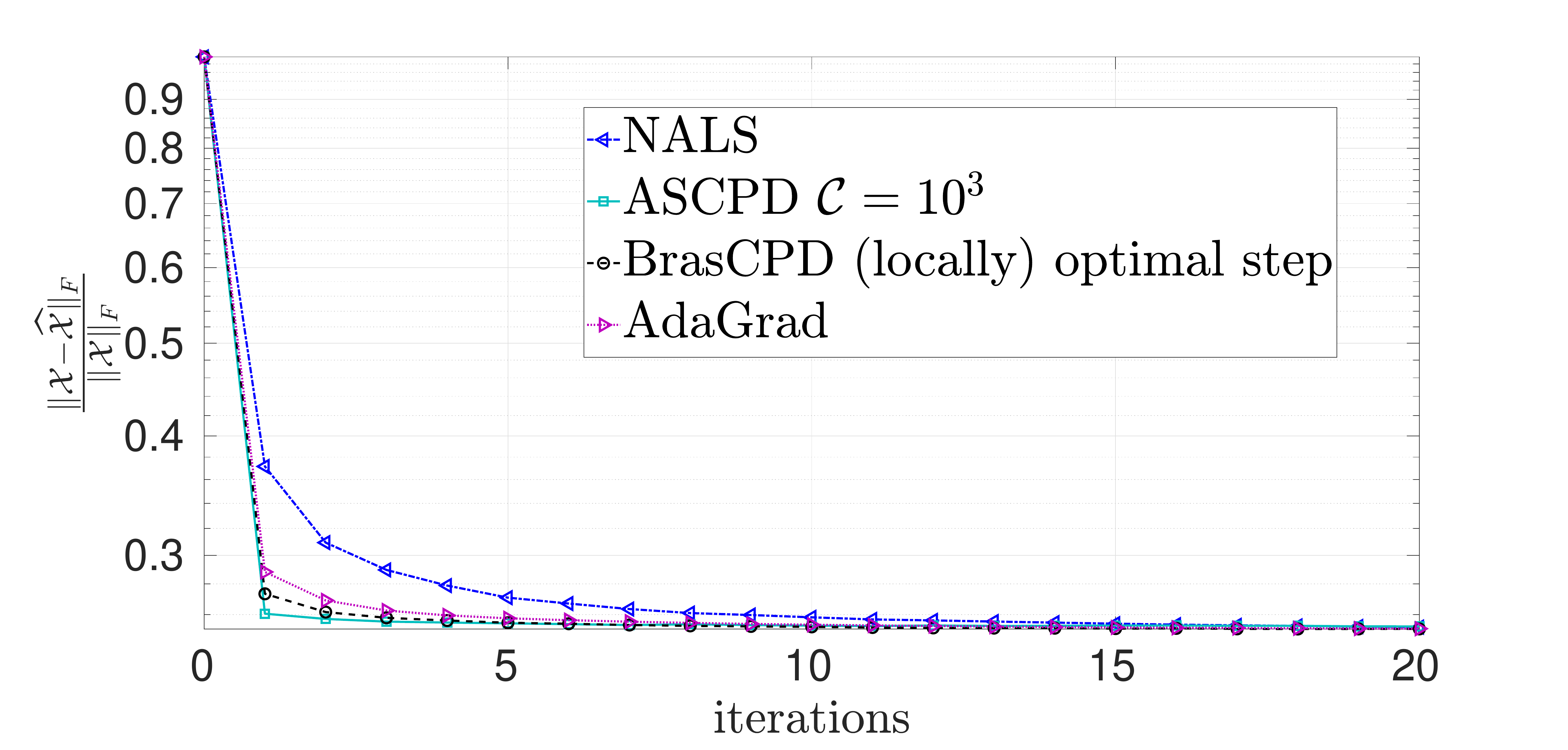}}
\centerline{\includegraphics[width=9.5cm]{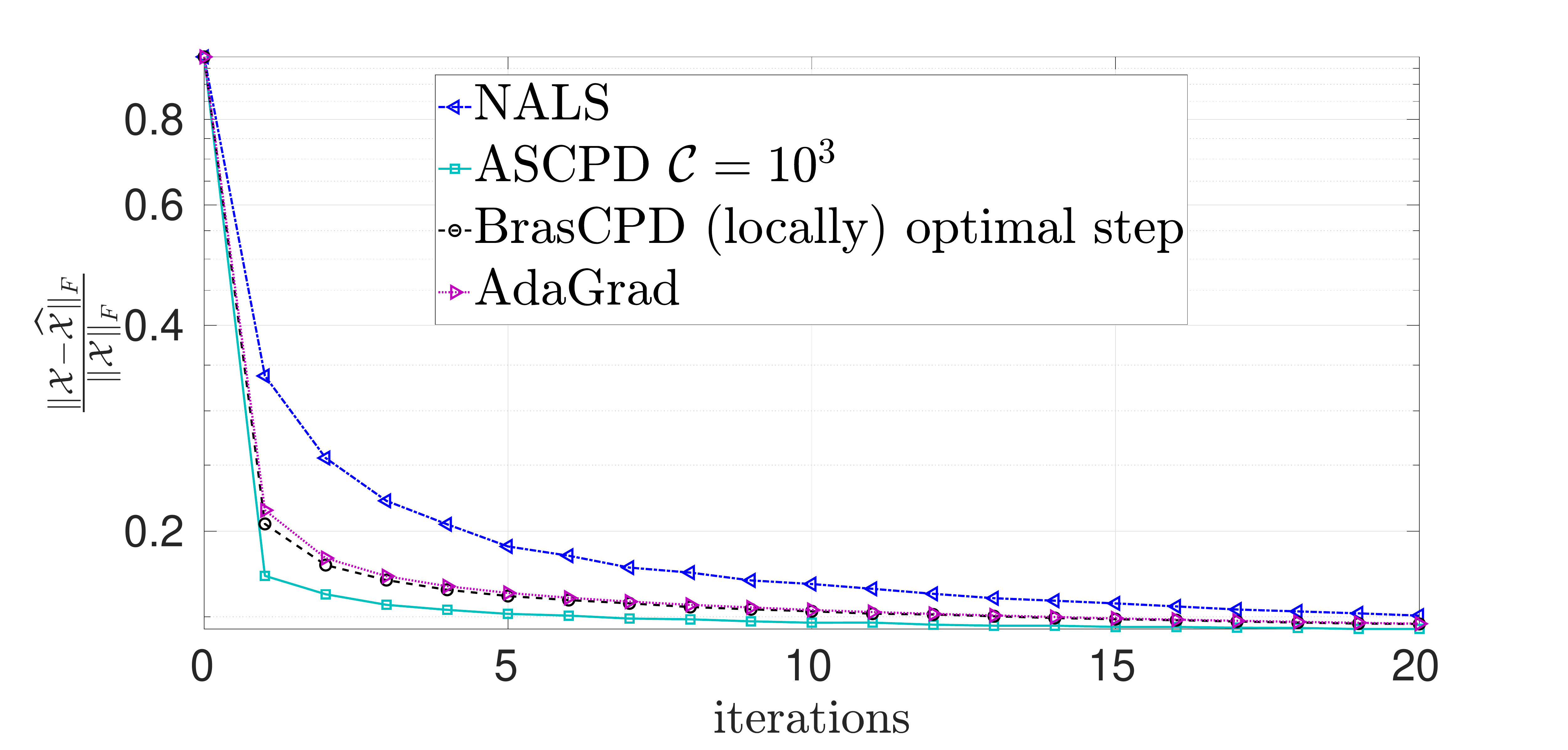}}
\caption{PaviaU dataset: relative tensor reconstruction error for
$B=500$ and $R=50$ (top), $R = 200$ (bottom).}
\label{fig_PaviaU}
\end{figure}

\subsection{Real-world Data}
\label{subsec:real}

Similarly to \cite{fu2020block}, we use the Indian Pine and PaviaU datasets which are Hyperspectral Images (HSIs). HSI sensors collect data in a group of images, for different wavelength ranges. The resulting data-cube is a third order tensor. The Indian Pine dataset is of size $145 \times 145 \times 220$ and consists of data acquired via the AVIRIS sensor, on Indian Pines site in Indiana (USA). The PaviaU dataset has size $610 \times 340 \times 103$ and consists of a scene of Pavia University in 
Italy.$^1$\begin{footnote*}[hb!]
{\small $^1$Datasets available at \url{http://www.ehu.eus/ccwintco/index.php/Hyperspectral_Remote_Sensing_Scenes}.
}
\end{footnote*}

In Fig. \ref{fig_indianpine}, we plot the quantity $m_k$ versus the number of
full iterations for the Indian Pine dataset for $B=500$ and $R=10$ (top) and $R=100$ (bottom).  

In Fig. \ref{fig_PaviaU}, we plot the corresponding results for the PaviaU dataset for 
$B=500$ and $R=50$ (top) and $R=200$ (bottom).  In all cases, we set $\eta=2$ and ${\cal C}=10^3$. 

We observe that, in all cases, the ASCPD outperforms all other methods in terms of $m_k$. 

\section{Conclusion}
\label{sec:conclusions}

We adopted a stochastic gradient based framework for the solution of the structured CPD problem. 
We improved upon known stochastic proximal gradient schemes by incorporating 
Nesterov-type acceleration using parameters that are ``locally optimal.'' 
In numerical experiments, with both synthetic and real-world data, our algorithm has been
proven efficient. Convergence analysis of the proposed algorithm is an interesting
future topic. 

\balance

\bibliographystyle{IEEEtran}


\end{document}